\documentstyle[12pt,epsfig]{article}
\let\section=\subsection     \let\subsection=\subsubsection                %%
\begin{document}
\begin{center}
   {\large \bf SUBTHRESHOLD $\rho^0$ PHOTOPRODUCTION}\\[2mm]
   {\large \bf ON $^2H,^3He$ and $^{12}C$ 
    \footnote{Proceedings of the International Workshop XXVIII
      on Gross Properties of Nuclei and Nuclear Excitations,
      Hirschegg, Austria, January 16-22, 2000, p. 203.}}\\[5mm]
   G.J. LOLOS \\[5mm]
   {\small \it  Department of Physics, University of Regina, Regina, SK, S4S 0A2,  
   Canada \\[8mm] }
\end{center}

\begin{abstract}\noindent
  The $\rho^0$ has been photoproduced using tagged photon energies in
  the 700-1120 MeV region on $\rm ^2H,^3He$ and $\rm ^{12}C$ by
  utilizing the Fermi momentum of the bound nucleons to produce the
  $\rho^0$. These energies lie mostly below the production threshold
  on the free proton.  Large mass modifications are indicated from the
  analysis, together with large polarizations of the produced $\rho^0$
  mesons with helicity zero, unlike the case of coherent $\rho^0$
  production on nuclei.
\end{abstract}

\section{Introduction}

The effects of hadron density and/or temperature on the mass and
lifetimes of vector mesons have been the subject of numerous
investigations, experimental as well as theoretical \cite{Ko97}.
Under conditions of chiral symmetry restoration, the quark condensate
vanishes, $<\bar{q}q> \rightarrow 0$. The condensate is not, however,
an observable but it is related to the meson mass, which is.  Thus,
the behaviour of vector meson masses under high hadron densities
and/or temparatures has become a testbed of our understanding of
hadron or quark dynamics in nuclear matter.  A renormalization of
vector meson masses in hot and dense matter is of significant
importance to the pursuit of quark gluon plasma research, has
relevance to the equation of state for nuclear matter in supernovae
explosions, and could affect lower energy nuclear physics, as well.
It has been argued that the mass modification of vector mesons should
be measurable even at normal nuclear matter densities \cite{Br91}.

Even though a reduction of vector meson mass in hadronic matter is
supported by most of the more recent theoretical models, the spectral
shape (manifested in the width) is more of an open question.  If there
is substantial broadening as reported in \cite{Kl97}, then extracting
meaningful mass change information from $\rho$ production and decay
experiments becomes problematic, especially so for massive nuclei.  In
this respect, production of either the $\omega$ in massive nuclei or
the $\rho$ from light nuclei may be a better solution.  In any case,
production of vector mesons, by whatever method, is not enough to
probe medium modifications.  The production of mesons must be
accompanied by a substantial decay fraction within the nuclear volume,
a fact which favours either low beam energies or restricted phase
space in which the vector mesons are essentially at rest with respect
to the recoiling nuclear system.

\section{The TAGX Results from $\bf A(\gamma_{t},\pi^+ \pi^-)X$ Reactions}

\subsection{The Subtreshold Energy Regime}

Photo- or lepto-production of vector mesons from nuclei has been and
is being explored at many facilities.  Given the quantum numbers of
the vector mesons and those of the photon (real or virtual), the
primary interaction of the latter proceeds via the interaction of the
former with the nucleons in nuclei, the well established VMD model of
$\gamma$-N interactions.  At incident photon energies of approximately
1.2 GeV and higher, the diffractive mechanism dominates, in which the
photon flactuates into a $\rho^0$ or $\omega$, which then scatters off
the nucleon field with sufficient four-momentum transfer to bring the
meson on the mass shell.  In the case of nuclei, the single $\gamma$-N
amplitudes add coherently and large production cross sections result.
Since the momentum transfer $t$ is less than 0.1 $\rm GeV^2/c^2$,
coherent or diffractive cross sections are very forward peaked.  As a
result, they also conserve helicity and the reaction is characterized
by a large impact parameter of approximately 10 fm or more.

While coherent photoproduction produces vector mesons prolifically,
the probability of the vector mesons penetrating the nuclear interior
and decaying within it is rather small.  The observation, then, of
medium modifications in such reactions is diminished by ``geometry''
(including trajectory of the produced meson with respect to the
nucleus) and the large Lorentz boost of the meson with respect to the
nucleus.  This is particularly severe for the $\omega$ due to its long
$c\tau$ compared to that of the $\rho^0$.  One is thus forced to
depend on the large number of mesons produced to observe the small
fraction that penetrates and decays within the nuclear volume, and
which is perharps overwhelmed in the invariant mass signature from the
low tail of the unmodified mesons decaying freely. Where the fraction
is larger, like in the $\rho$-meson case, the free $\rho^0$ width is
also large, thus effectively masking the modified mass. In the case of
the $\omega$, the narrow width makes it easier to separate the
modified from the unmodified masses, but the fraction of the former to
the latter is very small.

An alternative is to produce the $\rho^0$ below the free $\rm \gamma +
N \rightarrow \rho^0$ nominal threshold of 1083 MeV.  In this case,
the reaction is characterized by large $t$ and the Fermi four-momentum
of the struck nucleon is utilized to bring the $\rho^0$ on the mass
shell.  A penalty one pays for such {\em subthreshold} $\rho^0$
production is the small cross section compared to coherent production.
There are numerous advantages, however.  First, the $\rho^0$
production vertex is forced to be within the nuclear volume (small
impact parameter) by the interaction of the photon with a bound
nucleon. The reaction then is quasi-free $\gamma$-N interaction in
nature.  Second, the $\rho^0$ is produced with small relative momentum
and Lorentz boost with respect to the struck nucleon or the nucleus,
thus maximizing the probability of decay within the hadronic volume.
Third, the large opening angle (of $\rm 180^o$ in the $\rho^0$ rest
frame) is essentially preserved in the lab frame due to the small
Lorentz boost and this provides a characteristic signature of the
$\rho^0$ decay, which can be used effectively in the suppression of
competing processes.

\subsection{The TAGX Results on $\bf ^3He$}

The choice of $^3He$ was made based on several reasons. (a): Its core
nuclear density is near saturation density for heavier nuclei, while
the small number of nucleons reduces final state interactions (FSI)
between the pions and the recoiling nucleus or nucleons. (b): The
kinematical analysis is much simpler due to the small number of
spectator nucleons and the Monte Carlo (MC) simulations carry more
confidence. And, (c): extrapolations of cross sections for background
processes on proton and $^2H$ are more accurate for $^3He$ than more
complex nuclei.  Thus, it was felt that any mass modification, even if
small, could carry higher confidence due to the simplicity of the
probe-target system.

The analysis was based on a detailed and kinematically constrained fit
to several distributions simultaneously for all processes that can
lead to a $\pi^+ \pi^-$ final state.  Details are published in
\cite{Lo98}, \cite{Hu98}, and \cite{Ka99}. The results were consistent
with a large mass modification for the $\rho^0$, which was also found
to depend on incident photon energy.  At photon energies as low as
$660\pm 40$ MeV, the best fit was obtained for a $m^*_{\rho^0} =
490\pm 40$ $\rm MeV/c^2$, while for $840\pm 40$ MeV the mass of the
$\rho^0$ was $m^*_{\rho^0} = 640\pm 40$ $\rm MeV/c^2$.  For the
highest energy bin available to the experiment (which was centered
around 1080 MeV), the data were shown to be insensitive to any
$\rho^0$ mass modification.  The conclusions survived a number of
different analyses, and even the data sets for two of the analyses
were different and independent of each other.

The analysis was based on the simultaneous fitting of a number of
``key'' variables, such as missing momentum ($p_{miss}$), missing mass
($m_{miss}$), invariant mass ($m_{\pi \pi}$), pion opening angle
($\theta_{\pi \pi}$), and emission angle of the di-pion system
($\theta_{IM}$).  All these are kinematical observables and follow
phase space distributions, applicable to each individual reaction
assumed.  For the 80~MeV wide photon energy bins in the analysis, the
assumption was made that the matrix element is constant.  The yield
then is the product of several factors:

\begin{equation}
Y = |M^*(s,t)|^2 \cdot \Phi_{^3He}(p) \cdot \Psi (s,I,J)
 \cdot [MBPS*F(m,\Gamma)]
\end{equation}

where Y = Yield, $|M^*|$ = invariant matrix element for the reaction,
$\Phi$ = the single nucleon momentum distribution, $\Psi$ = function
of spin, isospin, and total angular momentum, and $[MBPS*F]$ = the
multi-body phase space for this reaction with the mass and width of
any resonances folded in.  Therefore, even though a variation of
$|M^*|$ within the 80 MeV bin is not taken explicitly into account,
the width $\Gamma$ of the states involved has a large part of this
energy dependence folded in.  Each background and foreground reaction
was simulated using Monte Carlo (MC) generators, which incorporated
equation 1.

Cross sections for known processes, such as $\Delta \pi,N^* \pi,\Delta
\Delta$ and $\pi^+ \pi^- \pi^0$, were extracted and compared to
established measurements in the literature, if such existed.  The
cross sections for $\rho^0$ production with modified masses were quite
small (a few $\mu$b), while that of competing processes are of the
order of 200~$\mu$b \cite{Hu98}.  The small $\rho^0$ signal could be
extracted due to the features of the TAGX acceptance.  The trigger
requirement of left-right (with respect to the photon beam) $\pi^+
-\pi^-$ coincidences, and the small out of plane acceptance of the
spectrometer, result in substantially larger acceptance for the
$\rho^0 \rightarrow \pi^+ \pi^-$ decay than the other competing
two-step processes leading to two-pion emission.  Furthermore, the
requirement of simultaneous fitting of so many different kinematical
variables provided severe contraints to the strength of these
background reactions in different regions of the various
distributions.

One of the difficulties of the above analysis is the model dependency
and the smallness of the $\rho^0$ production cross sections compared
to all the other competing processes.  The analysis, however, has been
shown to be insensitive to reasonable variations in assumed cross
sections for the dominant background processes because the $\rho^0$
events occupy quite distinct regions of phase space.  Thus, variations
in $\Delta \pi$ production cross sections, for example, affected
$\Delta \Delta$ fits and resulting cross sections, but they did not
affect the value of $m^*_{\rho^0}$.  Another conclusion was that
$m_{\pi \pi}$ is not a very sensitive parameter and good fits can be
obtained for large differences in the input parameters.  On the other
hand, $\theta_{\pi\pi}$ and $p_{miss}$ are difficult (and sensitive)
distributions to fit.

In the case of coherent $\rho^0$ production and at very small angles
of emission in the helicity frame of reference, the angular
distribution for $\theta^*_{\pi}$ of one pion with respect to the
di-pion center of mass momentum in the helicity frame, has a
$\sin{\theta^*_{\pi}}$ -like distribution for unpolarized photons.
This is a result of $s$-channel helicity conservation.  In the case of
helicity non-conservation, for example at large angles of $\rho^0$
emission due to large $t$ exchange, a random distribution of
polarization would result in an isotropic distribution for
$\theta^*_{\pi}$.  When, however, such a $\theta^*_{\pi^+}$
distribution was formed, an enhancement in the regions of $0^o$ and
$180^o$ was observed in the ``raw'' event sample.  This has been
interpreted as a substantial polarization component in the $\rho^0$
event sample which corresponds to the $l=1, \; m=0$ substate.  Such a
$Y^0_1$ -like distribution was most prominent in the same $m_{\pi\pi}$
bins where the independent kinematical analysis placed the mean of the
modified $\rho^0$ mass.

\subsection{The TAGX Results on $\bf ^2H, \; ^{12}C$}
 
The analysis of the $\rm ^2H$ and $\rm ^{12}C$ results is still
ongoing and a final value for the modified masses has not been
extracted yet.  The philosophy and methodology is quite different than
the analysis of the $\rm ^3He$ data.  The simpler $\rm ^2H$ nucleus,
coupled with photoabsorption cross sections in the literature for most
of the channels leading to di-pion emission, allow the use of
experimental amplitudes and cross sections and a likelyhood method of
fitting the kinematical observables.  Unlike the analysis of the $\rm
^3He$ data, in this analysis the $\theta^*_{\pi^+}$ distributions are
explicitly involved among the kinematical variables fitted.  The cross
section results are not yet final, however, the analysis based on the
$\theta^*_{\pi^+}$ distribution is quite interesting.  In Figure 1,
the $\cos{\theta^*_{\pi^+}}$ distributions are shown for $\rm ^{12}C
+ ^2H$ data, (a) without any restrictions on the data, and (b) with a
$\theta_{\pi\pi} \ge 120^o$ cut.  Several observations can be drawn:

\begin{itemize}
\item {In the $M_{\pi\pi}$ bin of 300-400 $\rm MeV/c^2$, the shape of
    the distribution and the small number of events near the -1 and +1
    limits is characteristic of acceptance effects due to trigger and
    geometry of TAGX.  As such, the population in these two regions
    for the invariant mass bins in the 500-600 $\rm MeV/c^2$ regions,
    signify substantial $\rho^0 \rightarrow \pi^+ \pi^-$ -like
    production.}
\item {The application of the opening angle ($\theta_{\pi\pi}$) cut,
    removes events form the central regions of the distribution, but
    leaves the population around the two limits unchanged.}
\end{itemize}

\begin{figure}[t]
\begin{center}
\epsfig{file=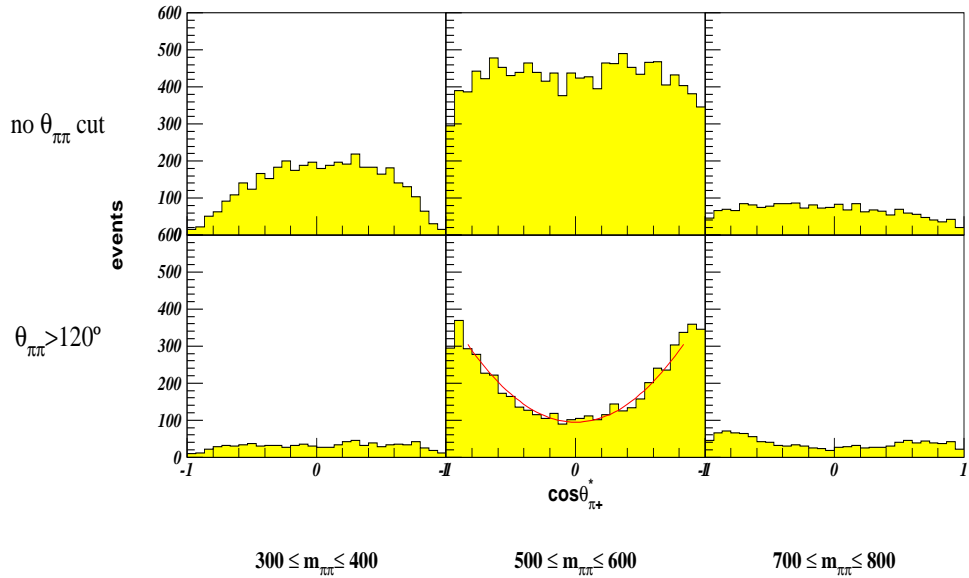,width=5.0in,height=3.0in,angle=0}
\caption{Selected invariant mass bins for data from the
  $\rm ^2H,^{12}C(\gamma,\pi^+ \pi^-)X$ experiment obtained with TAGX.
  The top panels have no $\theta_{\pi \pi}$ cut, whereas the bottom
  panels have been subjected to a $\theta_{\pi \pi} \ge 120^o$ cut.
  The invariant mass bins for 400-500 and 600-700 MeV/c$^2$ also
  exhibit a $p$-wave signature, but it's neither as strong nor as
  symmetric as that in the 500-600 MeV/c$^2$ bin; they are not shown
  here for the sake of clarity.}
\label{fig:thetacut}
\end{center}
\end{figure}

The $\cos{\theta^*_{\pi^+}}$ distributions for all three nuclei
exhibite the same overall behaviour.  The preliminary results from the
cross section fitting of the $\rm ^2H$ data are consistent with a few
$\mu$b cross section for polarized $\rho^0$ production in the $l=1, \;
m=0$ substate and no strength in the $l=1, \; m=1$ substate.  This is
qualitatively in agreement with the $\rm ^3He$ data which also showed
no statistically significant population in the small angle regions in
plots such as shown in Figure~\ref{fig:thetacut}.  This was verified
by successively applying increasingly tighter opening angle cuts and
comparing the behaviour as a function of $\cos{\theta^*_{\pi^+}}$ values.

\subsection{The Effects of $\bf \theta_{\pi\pi}$ Cuts} 

\begin{figure}[t]
\begin{center}
\epsfig{file=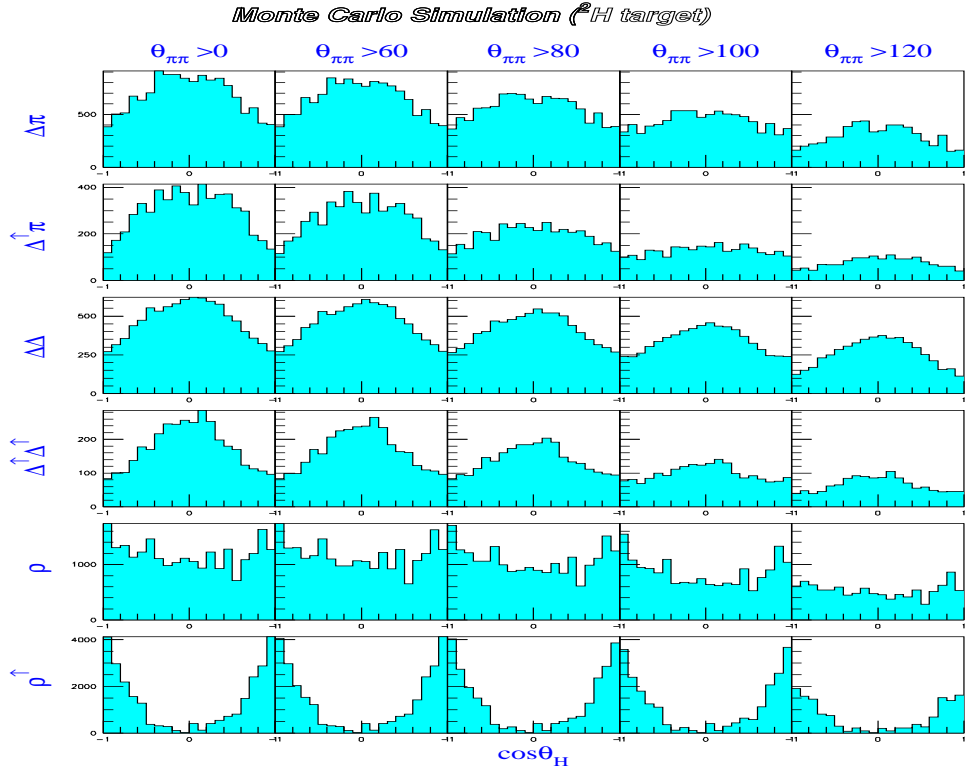,width=5.0in,height=4.0in,angle=0}
\caption{Deuterium distributions of the cosine of the pion angle 
  in the helicity frame.  The columns indicate the effect of
  progressively larger opening angle cut.  The rows portray the
  different MC-generated channels, with the verticle arrow indicating
  that the particle in question is produced in a polarized state.}
\label{fig:newpwave}
\end{center}
\end{figure}

The $l=1, \; m=0$ signature is model independent and difficult to
argue away.  Nevertheless, there have been a number of possible
explanations, other than a modified and polarized $\rho^0$.  One is
that {\em there may be} other reactions which lead to two-pion
emission in a relative $p$-state and with a $m=0$ substate.  However,
no such specific reaction has ever been identified or proposed as a
candidate; in any case, if there is indeed one, this by itself is a
new and interesting result.  Another concern put forward is that the
application of ever tighter $\theta_{\pi\pi}$ requirements may
artificially produce such $l=1, \; m=0$ signatures.  In order to
address this valid concern, a detailed investigation of the behaviour
of two of the most prolific background processes has been made as a
function of $\theta_{\pi\pi}$ cuts.  MC simulations of three
production reactions have been made and analyzed in the same fashion
as the real data are analyzed in the TAGX detector. These are shown in
Figure~\ref{fig:newpwave} for $\rm ^2H$, plotted as a function of
$\cos{\theta_H}$ in the helicity frame.

The $\Delta \pi$ and $\Delta \Delta$ background channels are the two
strongest channels leading to $\pi^+\pi^-$ production.  As can be seen
in Figure~\ref{fig:newpwave}, whether the $\Delta$ is produced
polarized (indicated by the arrow) or not, the application of
successively more demanding $\theta_{\pi\pi}$ cuts does not alter the
shape of the distributions.  In the case of unpolarized $\rho^0$
production, the distribution is essentially flat with a slight
enhancement at the limits due to detector acceptance.  Only in the
case of polarized $\rho^0$ production do the MC simulations in
Figure~\ref{fig:newpwave} agree with the observed behaviour of the
data in Figure~\ref{fig:thetacut}.

\section{Conclusions}

The analyses of $\rm ^3He$ data using MC and model dependent, as well
as model independent variables, agree on a substantial mass
modification of the $\rho^0$. The preliminary results on $\rm ^2H$ and
$\rm ^{12}C$, using primarily the model independent $\theta^*_{\pi^+}$
analysis, agree with the earlier results and strongly support the
production of longitudinally polarized (zero helicity) and mass modified
$\rho^0$ -mesons. The similarities among the results for these three
different nuclei is suggestive of a $\rho^0$-N based interaction
effect rather than a nuclear effect.  Since the relative $\rho^0$-N
momenta at such low incident photon energies are also very low, the
mean decay length of the $\rho^0$ from its production vertex is less
than the nucleon radius.  Hadronic densities at the $interior$ of a
nucleon are very high, thus qualitatively accounting for the large
mass modifications reported so far from TAGX results. The excellent
agreement on the $\theta_{\pi\pi}$ dependence for data and MC
simulations conclusively shows that only a $l=1, \; m=0$ substate of
the $\rho^0$ is present in the data.  Such a production can be
accounted for by a spin-flip transition for the target nucleon,
consistent with a $N^*$ excitation in the $\rho^0$ production process.

\end{document}